\RequirePackage{lineno}
\documentclass[a4paper]{jpconf}

\usepackage{graphicx}
\usepackage{amsmath}
\usepackage{caption}
\usepackage{subcaption}

\begin{document}

\title{Fully hadronic $t\overline{t}$ cross section with the ATLAS detector}

\author{Andrea Coccaro on behalf of the ATLAS Collaboration}

\address{Department of Physics, University of Washington, Seattle WA, United States of America}

\ead{Andrea.Coccaro@cern.ch}

\begin{abstract}
A measurement of the $t\overline{t}$ production cross section in the all-hadronic decay mode is presented using 4.7~fb$^{-1}$ of proton-proton collisions at a centre of mass energy of $\sqrt{s}=7$ TeV collected by the ATLAS experiment in 2011. Events are selected using a multi-jet trigger. Kinematic and $b$-tagging requirements are then applied to identify $t\overline{t}$ event candidates. A kinematic fit reconstructs the event topology of the final state extracting the top-quark mass which is then used to measure the production cross section with an unbinned likelihood fit. The result is found in good agreement with the Standard Model prediction for a top-quark mass of 172.5~GeV.
\end{abstract}

\section{Introduction}
The measurement of the top-quark pair production cross section provides an important test of perturbative QCD calculations and top-quark pair production constitutes a major background source to many new physics scenarios predicted to be accessible at the LHC. The $t\overline{t}$ production cross section is nowadays calculated at approximate NNLO and is $\sigma_{t\overline{t}}=167^{+17}_{-18}$~pb for a top-quark mass of 172.5~GeV at a centre-of-mass energy of $\sqrt{s}=7$~TeV \cite{TheoryPaper}. 

In the fully hadronic final state, the two $W$'s decay hadronically and the experimental signature is characterized by a nominal six-jet topology with two jets stemming from the hadronization of $b$-quarks. The all-hadronic channel has the advantage of a large branching ratio, 46\%, although it suffers from a huge background of multi-jet events.

The present result \cite{Analysis} is based on data recorded during the 2011 run with $pp$ collisions at $\sqrt{s}=7$~TeV using the ATLAS detector \cite{DetectorPaper}. The total integrated luminosity is 4.7~fb$^{-1}$ with an uncertainty of 3.9$\%$ after asking for stable beam conditions and all subsystems to be operational. This analysis is complementary to similar measurements in the single lepton and dilepton channels. %It is combined using the ATLAS cross-section measurements in the other top-quark decay modes to derive an average value of the $t\overline{t}$ production cross section at the LHC centre-of-mass energy.

\section{Trigger, event selection and simulated samples}\label{sec:EventSelection}
The events considered in this analysis are triggered asking for at least five jets with $|\eta|<3.2$ and $E_T>30$~GeV. The calibration constants to correct for the lower hadron response of the non-compensating calorimeters were not included in the trigger system and an offline cut of $E_T>55$~GeV is introduced at the analysis level to operate nearby the plateau of the trigger selection. The 55~GeV cut is obtained by studying the trigger dependence on the fifth leading jet and corresponds to a 90$\%$ efficiency with respect to jets reconstructed offline. The efficiency plateaus at more than 99$\%$ starting from 60~GeV and 55~GeV represents the best compromise between signal acceptance and contribution to the trigger systematic uncertainty.
%The efficiency is found to be at the plateau at 60~GeV and 55~GeV represents the best compromise between signal acceptance and contribution to the trigger systematic uncertainty.

To select a sample enriched in top-quark pair events, a preliminary set of cuts is applied on the triggered events: at least one reconstructed primary vertex with at least five associated tracks, no isolated leptons with $p_T>20$~GeV, at least five jets with $E_T>55$~GeV and $|\eta|<2.5$ and at least one more jet with $E_T>30$~GeV and $|\eta|<2.5$. At least two jets, out of the five leading jets, are required to be $b$-tagged combining the high-performance algorithms \cite{bTagginaATLAS} in a neural network. Jets with an unlikely association to the hard scatter in the event, using the jet vertex fraction information \cite{JVFCut}, are not considered in the analysis. 

The preselection is completed requiring a minimum angular distance between the two $b$-tagged jets of $\Delta R>1.2$ and a minimum angular distance between any two reconstructed jet of $\Delta R>0.6$. The former cut reduces the background contamination from gluon splitting without compromising the signal efficiency while the latter cut is needed to avoid relying on the trigger simulation in case of close-by jets.

The modelling of the $t\overline{t}$ signal and its associated selection efficiency is derived from MC simulation. The MC@NLO generator with PDF set CT10 is used for the generation of the $t\overline{t}$ signal, assuming a top quark mass of 172.5~GeV. Due to the large uncertainty in the multi-jet cross-section prediction, a data-driven technique is employed to estimate the background, while the MC simulation is only used to estimate the uncertainty on the mass shape distribution for the background.

%\begin{center}
%\begin{figure}[th!]
%\begin{minipage}[b]{0.5\textwidth}
%\includegraphics[width=15pc]{./Figures/fig_01a.png}
%\end{minipage}
%\begin{minipage}[b]{0.5\textwidth}
%\includegraphics[width=15pc]{./Figures/fig_01b.png}
%\end{minipage}
%\caption{}
%\end{figure}
%\end{center}
\begin{figure}[t!]
\centering
\begin{subfigure}[b]{0.48\textwidth}
\includegraphics[width=15pc]{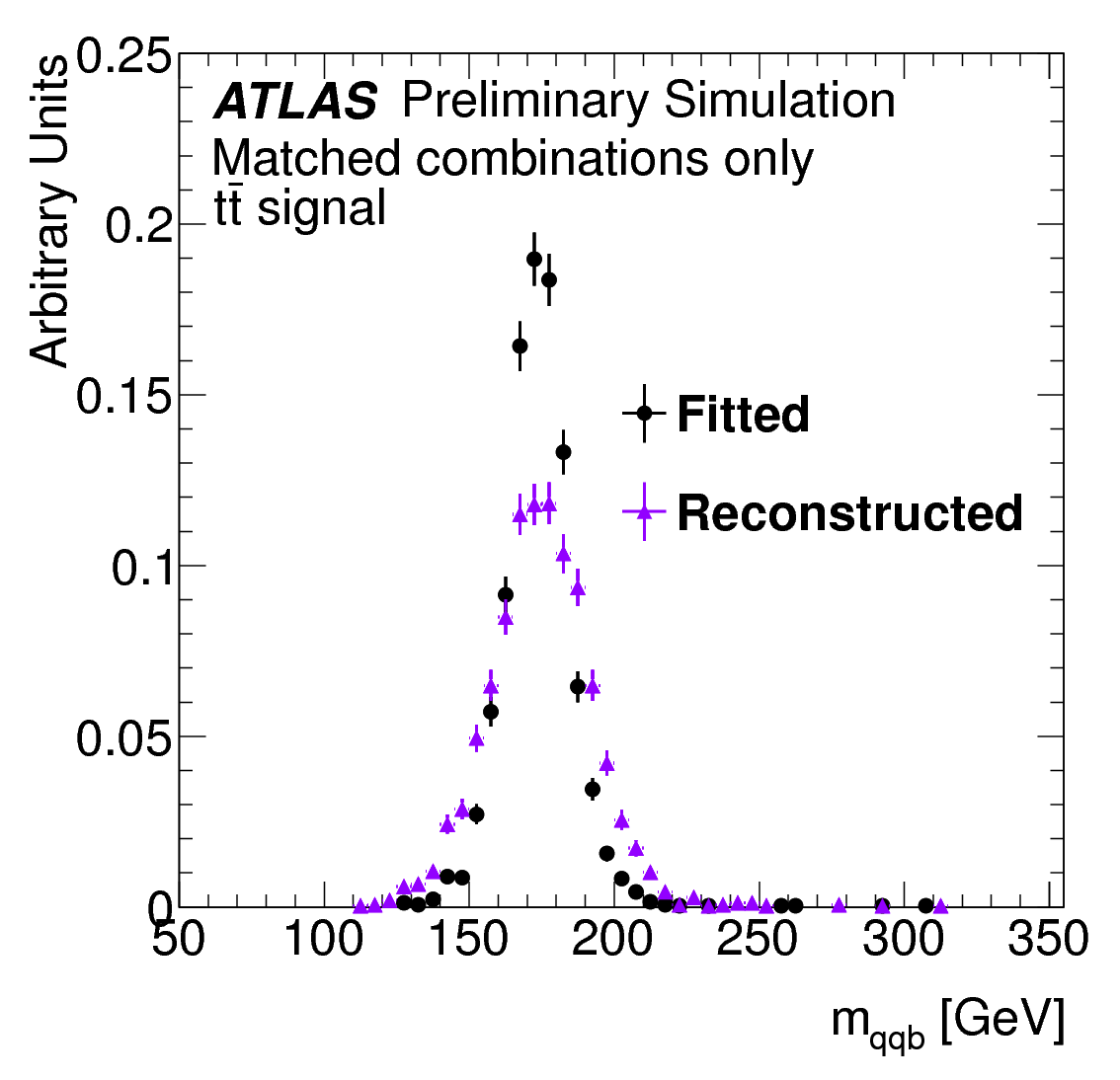}
\caption{(a)}
\end{subfigure}
\begin{subfigure}[b]{0.48\textwidth}
\includegraphics[width=15pc]{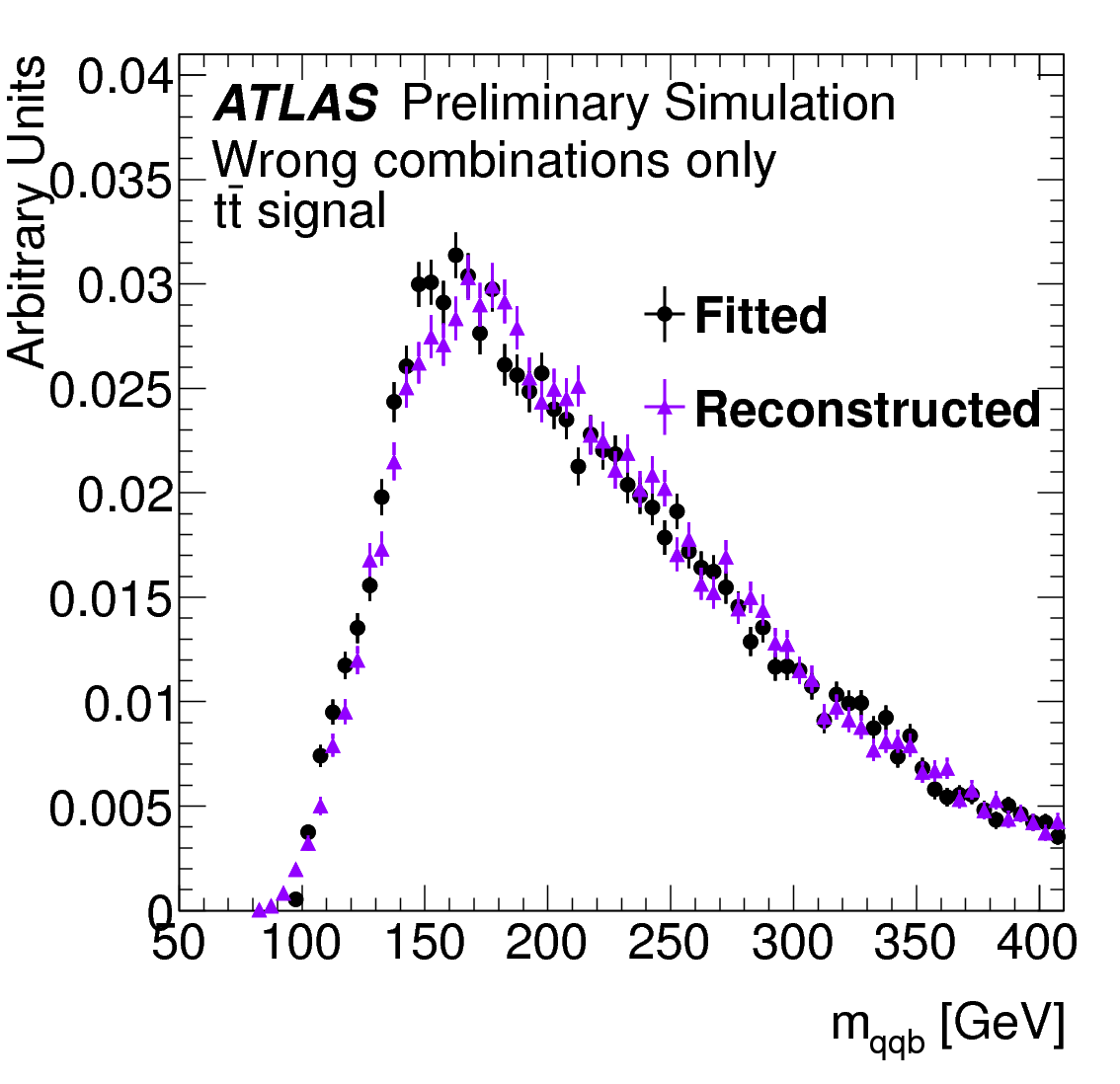}
\caption{(b)}
\end{subfigure}
\caption{Top-quark mass distribution using the reconstructed or the corrected jet energies after the kinematic fit described in the text \cite{Analysis}. Only events where all the jets are correctly matched to quarks are shown in (a). Events where at least one jet is wrongly matched are shown in (b).}
\label{fig:TopMassDistribution1}
\end{figure}

\section{Kinematic fit}
A kinematic fit is performed on the selected events to optimize the top-quark mass reconstruction. The fit is based on a likelihood approach to find the correct association of jets with the final state partons of the fully-hadronic $t\overline{t}$ decays. The likelihood function $\mathcal{L}$ includes Breit-Wigner functions used to constrain the di-jet and the tri-jet masses to the $W$-boson mass and the top-quark mass, respectively. The top-quark mass is treated as a free parameter but it is constrained to be identical for the top and anti-top quark candidates. The $W$-boson mass and width are fixed to the known values in literature. Additional terms are added taking into account the $b$-tagging algorithm response biasing the likelihood to select permutations in which a $b$-tagged jet is matched to a $b$-quark.

The fit is performed for all possible permutations of jets and quarks and for each permutation the function $-\ln{\mathcal{L}}$ is minimized with respect to the six jet energies and the reconstructed top-quark mass. The sum of all permutations is computed and the relative weight, normalized to unity, of the best computation is referred to as the {\it event probability}. 

The method is tested on simulated events passing the selection cuts. Figure \ref{fig:TopMassDistribution1}\,(a) shows the mass distribution obtained for the events where all the jets are correctly matched to the corresponding quarks using the MC truth information. As expected, the mass resolution is improved when using the jet energies calculated during the kinematic fitting procedure compared to just using the reconstructed jet energies. The fraction of signal events fully matched by the kinematical fit is found to be 11\%. For the events where at least one jet is wrongly matched there is no gain in the mass resolution by performing the kinematic fit, as shown in Figure \ref{fig:TopMassDistribution1}\,(b).

\section{Cross section measurement}
Few additional cuts are imposed before measuring the $t\overline{t}$ production cross section. Events are required to have $m_t>125$~GeV and a jet multiplicity between 6 and 10. To further improve the signal to background ratio, cuts are introduced requiring the {\it event probability} to be greater than $0.8$ and the minimal mass $\chi^2$ to be less than $30$. The $\chi^2$ is defined as:
\begin{equation*}
\chi^2=\frac{(m_{j_1,j_2}-m_W)^2}{\sigma^2_W}+\frac{(m_{j_1,j_2,b_1}-m_t)^2}{\sigma^2_t}+\frac{(m_{j_3,j_4}-m_W)^2}{\sigma^2_W}+\frac{(m_{j_3,j_4,b_2}-m_t)^2}{\sigma^2_t}\,,
\end{equation*}
where $j_1, j_2, j_3, j_4, b_1$ and $b_2$ are the reconstructed jets and $b$-jets and $m_W, m_t, \sigma_W$ and $\sigma_t$ are obtained by the mean and the half-width gaussian parametrization of the top-quark and $W$-boson mass distributions given by the all-hadronic $t\overline{t}$ MC simulation. The minimal mass $\chi^2$ is defined as the minimum $\chi^2$ obtained when computing its value for every permutation per event. With this final event selection, the fraction of signal events fully matched by the kinematical fit rises to 36\%.

\begin{figure}[t!]
\centering
\begin{subfigure}[b]{0.48\textwidth}
\includegraphics[width=15pc]{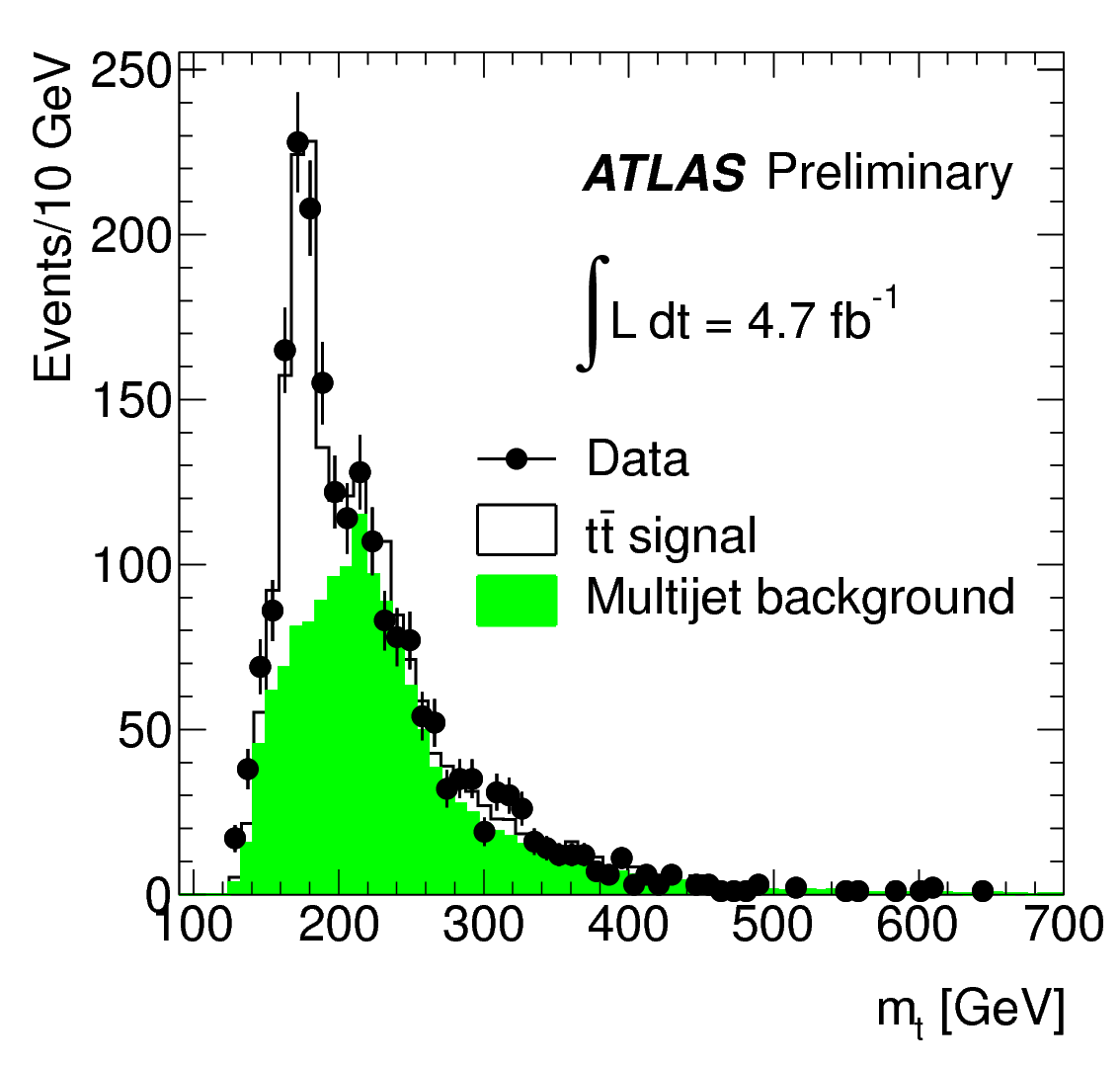}
\caption{(a)}
\end{subfigure}
\begin{subfigure}[b]{0.48\textwidth}
\includegraphics[width=15pc]{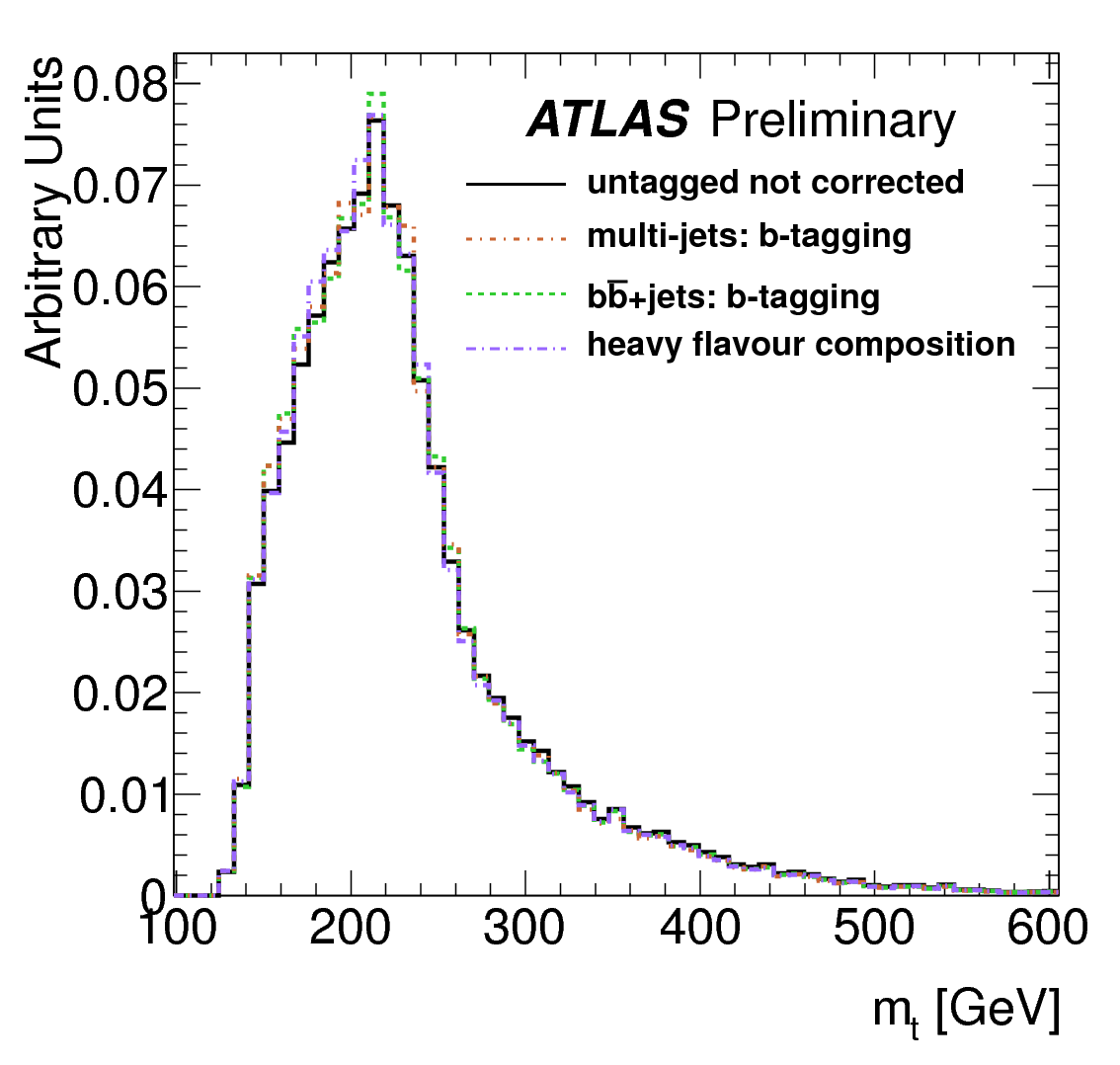}
\caption{(b)}
\end{subfigure}
\caption{(a) Unbinned likelihood fit of the top-quark mass distribution \cite{Analysis}. The signal and background templates are also displayed and the error bars associated to the data are statistical only. (b) Top-quark mass distribution after the kinematic fit using the sample selected with all the cuts described in the text but dropping the $b$-tagging requirement \cite{Analysis}. This distribution is used as a template of the multi-jet background for the likelihood fit and it is also shown in Figure \ref{fig:TopMassDistribution2} (a). Additional distributions obtained applying different corrections derived from MC simulations are overlaid using dashed lines.}
\label{fig:TopMassDistribution2}
\end{figure}

An unbinned likelihood fit is performed on the final data sample, shown in Figure \ref{fig:TopMassDistribution2}\,(a). The signal template distribution is derived from MC simulation and includes both the $t\overline{t}$ combinatorial background and the correct combinations while the background template distribution is derived from data events in the same signal region but without imposing any $b$-tagging requirement. The signal fraction is estimated by the fitting procedure to be $31.4\pm2.3\%$ and considering the inclusive $t\overline{t}$ selection efficiency, derived from MC simulation, the total cross section is measured to be $\sigma_{t\overline{t}}=168\pm12\mbox{(stat.)}$~pb.

%\begin{center}
%\begin{figure}[th!]
%\begin{minipage}[b]{0.5\textwidth}
%\includegraphics[width=15pc]{./Figures/fig_02a.png}
%\end{minipage}
%\begin{minipage}[b]{0.5\textwidth}
%\includegraphics[width=15pc]{./Figures/fig_02b.png}
%\end{minipage}
%\label{fig:TopMassDistribution2}
%\end{figure}
%\end{center}

The events passing all the cuts but dropping the $b$-tagging requirement are found to be dominated by multi-jet events, with a residual fraction of all-hadronic $t\overline{t}$ events estimated to be 4.6\%. The top-quark mass distribution obtained with this methodology, referred as to the untagged sample, is shown in Figure \ref{fig:TopMassDistribution2}\,(b) and is used as a template for the background in the likelihood fit. Any $b$-tagging technique biases the jet $p_T$ distribution and therefore most of the kinematic distributions, such as the top-quark mass spectrum, are affected by systematic uncertainties related to the $b$-tagging requirement. To estimate the uncertainty on the background modelling, the top-quark mass shape is studied using simulated events for both generic multi-jet and exclusive $b\overline{b}+$jets events and compared with and without the $b$-tagging requirement. The ratio of the top-quark mass distributions between the tagged and untagged and between the multi-jet and $b\overline{b}+$jets samples are computed and applied to the untagged data sample. The effect of these corrections on the data-driven background is shown in Figure \ref{fig:TopMassDistribution2}\,(b) and the maximum variation between the nominal and the corrected cross sections gives a systematic uncertainty on the background modelling of 4\%.

The highest contribution to the total systematic uncertainty of the cross section measurement is from the jet energy scale and is estimated to be +20/-11\%. Other sources of uncertainty were estimated and are related to the $b$-tagging efficiency and mistag rate ($\pm17$\%), the initial and final state radiation ($\pm17$\%), the dependence on the generator and on the parton shower ($\pm13$\%), the multi-jet trigger efficiency ($\pm10$\%), the parton distribution function (+7/-5\%), the luminosity ($\pm3.9\%$) and the jet energy resolution ($\pm3$\%). The total systematic uncertainty is estimated to be +36/-34\%.

\section{Conclusions}
The production cross section of $t\overline{t}$ events in the all-hadronic decay channel was measured at the LHC using 4.7~fb$^{-1}$ of $pp$ collisions at the LHC with the ATLAS detector. The vastly dominant background consists of multi-jet events and its shape is modelled using a fully data-driven technique. A kinematic fit is employed to derive the top-quark mass distribution which is then used to extract the cross section measurement with an unbinned likelihood fit. The final measurement is $\sigma_{t\overline{t}}=168\pm12\mbox{\,(stat.)}^{+60}_{-57}\mbox{\,(syst.)}\pm7\mbox{\,(lum)}$~pb. The measured cross section is found to be in agreement with the theoretical expectation of $\sigma_{t\overline{t}}=167^{+17}_{-18}$\,pb \cite{TheoryPaper}.

\end{document}